\documentclass[conference]{IEEEtran}
\usepackage{lipsum}
\usepackage{fancyhdr}
\usepackage{url}
\usepackage{fancyheadings}
\fancypagestyle{firstpage}{%
  \fancyhf{}
  
  \fancyfoot[C]{\textit{\small{IEEE International Conference On Cyber Security And Protection Of Digital Services (Cyber Security 2017), June 19-20, 2017, London, UK.}}}
}

\ifCLASSINFOpdf
  \usepackage[pdftex]{graphicx}
\else
\fi
\begin{document}
%
\title{Improving Dynamic Analysis of Android Apps Using Hybrid Test Input Generation}

\author{\IEEEauthorblockN{Mohammed K. Alzaylaee, Suleiman Y. Yerima, Sakir Sezer}
\IEEEauthorblockA{Centre for Secure Information Technologies (CSIT)\\
Queen's University Belfast\\
Belfast, Northern Ireland\\
Email: \{malzaylaee01, s.yerima, s.sezer\}@qub.ac.uk}
}


%


\maketitle

\thispagestyle{firstpage}

\begin{abstract}
The Android OS has become the most popular mobile operating system leading to a significant increase in the spread of Android malware. Consequently, several static and dynamic analysis systems have been developed to detect Android malware. With dynamic analysis, efficient test input generation is needed in order to trigger the potential run-time malicious behaviours. Most existing dynamic analysis systems employ random-based input generation methods usually built using the Android Monkey tool. Random-based input generation has several shortcomings including limited code coverage, which motivates us to explore combining it with a state-based method in order to improve efficiency. Hence, in this paper, we present a novel hybrid test input generation approach designed to improve dynamic analysis on real devices. We implemented the hybrid system by integrating a random based tool (Monkey) with a state based tool (DroidBot) in order to improve code coverage and potentially uncover more malicious behaviours. The system is evaluated using 2,444 Android apps containing 1222 benign and 1222 malware samples from the Android malware genome project. Three scenarios, random only, state-based only, and our proposed hybrid approach were investigated to comparatively evaluate their performances. Our study shows that the hybrid approach significantly improved the amount of dynamic features extracted from both benign and malware samples over the state-based and commonly used random test input generation method. 
\end{abstract}



%
\IEEEpeerreviewmaketitle

\section{Introduction}
Smartphones are becoming an essential electronic device of everyone's daily life. With nearly 80\% market share, the Google Android Operating system (OS) is the leading OS in the market compared to iOS, Windows, Blackberry, and Symbian mobile. Over 65 billion downloads have been made from the official Google play store and there are currently more than 1 billion Android devices worldwide ~\cite{president}. Statista ~\cite{statista} reports there will be more than 1.5 billion Android devices shipped worldwide by 2020. At the same time, malware targeting Android devices has increased significantly over the last few years. According to a report from McAfee, there are around 2.5 million new Android malware samples exposed every year which increased the total number of malware samples discovered in the wild to more than 12 million ~\cite{McAfeeLabs2016}. Android malware can be found in a variety of applications such as gaming apps, banking apps, social media apps, educational apps, utility apps etc. Malware-infected apps may have access to privacy-sensitive information, send text messages to  premium rate numbers without user approval, or even install a rootkit on the device enabling it to download and execute any code the malware developer wants to deploy etc. 

To mitigate the spread of malware, Google introduced Bouncer to its Play store in Feb 2012. Bouncer is the system used to monitor submitted apps for potentially harmful behaviours. It uses a sandbox to test the submitted applications  for five minutes in order to spot any malicious activities. However, Bouncer can be evaded by means of some simple detection avoidance  techniques ~\cite{Oberheide2012}. Furthermore, most third party app stores do not have any screening mechanisms for submitted apps. There is therefore a need for efficient detection mechanisms to detect zero-day Android malware in the wild.  Several approaches for detecting Android malware have been proposed in previous works. These approaches are categorized into either static or dynamic analysis or utilize both.

In the static analysis approach, the code is usually reverse engineered and examined for presence of any malicious code. ~\cite{Eigenspace, Drebin, Yerima, DroidAPIMiner, reducing} are a few examples of  detection solutions based on static analysis. While in the dynamic analysis method, the application is executed in a controlled environment, such as a sandbox or virtual machine, or a physical device with the purpose of tracing its behaviour. Researchers have proposed several types of automated dynamic analysis systems to detect suspicious activities and behaviours from Android apps ~\cite{Enck2010,DroidBox, Tam2015,AppsPlayground,tracedroid, SandDroid, Alzaylaee2016}. However, the efficiency of these systems depend on the ability to effectively trigger the malicious behaviours during the analysis. 

Android apps are heavily user interface (UI) driven, which makes automated dynamic analysis on smartphones more difficult compared to traditional desktop scenarios. Because of the UI operation, efficient input generation is very important for testing apps and many tools are available to aid developers. Likewise, efficient input generation is needed to drive automated dynamic analysis for malware detection. On the Android platform, malware can hide their malicious activities behind events that require user interaction. Therefore, to test the apps dynamically, researchers need a tool that can simulate the human input to trigger these apps to start their malicious behaviours. The main goal is to reach a high percentage of  code coverage such that most of the suspicious activities are revealed during the analysis. Code coverage refers to the amount of the source code that has been traversed and executed during the dynamic analysis of the apps. Many of the existing dynamic analysis systems such as AppsPlayGround ~\cite{AppsPlayground}, A3E ~\cite{azim2013targeted}, and DynoDroid ~\cite{dynodroid} rely on a random-based input generation strategy based on the Android Monkey UI exerciser tool ~\cite{monkey}. According to an empirical study ~\cite{automated}, Monkey reached the highest code coverage compared to other test input generation methods. In particular, the study showed that Monkey achieved higher code coverage than more sophisticated test input generation systems like Dynodroid ~\cite{dynodroid}, ACTEve ~\cite{anand2012automated}, A3E ~\cite{azim2013targeted}, GUIRipper ~\cite{amalfitano2012using}, SwiftHand ~\cite{choi2013guided}, and PUMA ~\cite{puma}. Due to the randomness of the inputs generated by Monkey, it may reach certain points in the phone where it is not intended or desired during the dynamic analysis. For example, Monkey could generate inputs that may turn on airplane mode, turn off Wi-Fi, or even turn off the USB debugging mode (which is necessary for device based analysis), etc. All of these will affect the performance of the dynamic analysis tools.

Hence, in order to overcome the aforementioned limitations, we are motivated to explore combining the advantages of different types of test input generation methods to increase the percentage of code coverage as much as possible and potentially collect more dynamic features (API call logs and intents). In this study, we implemented a novel strategy by combining a random-based method (implemented using Monkey ~\cite{monkey}) with the state-based method (implemented using DroidBot ~\cite{droidbot}). Monkey is designed to provide  developers with a random testing tool. Whereas DroidBot is a tool designed to intelligently understand the app operation by statistically analysing its code and dynamically analysing the app User Interface (UI) hierarchy. We have implemented the hybrid input generation strategy using python and incorporated this into the DynaLog ~\cite{Alzaylaee2016} dynamic analysis tool. Furthermore, in order to validate this approach, we examine the performance of the hybrid method by comparative analysis of Monkey vs. Hybrid, DroidBot vs. Hybrid, Monkey vs. DroidBot, and Monkey vs. DroidBot vs. Hybrid in order to evaluate the performance in terms of number of dynamic API logs and system events collected during analysis. We speculate that the higher the code coverage the more the number of API related features we would be able to extract from the apps. The experiments were performed on real devices using 1222 benign samples from McAfee Labs (Intel Security) and 1222 malware samples from the Android malware genome project ~\cite{malgenomeproject}.

The reminder of the paper is structured as follows. Section II discusses the Android input generation/application triggering tools. Section III details the methodology and experiments undertaken to evaluate the hybrid approach. Section IV presents and discusses the results. Section V discusses the related work, followed by conclusions and future work in Section VI.

\section {Triggering Android Applications}

As mentioned earlier, there are a number of tools available for dynamic analysis of Android applications. As the main goal is to analyse and detect zero-day malware, there is a need to find the most effective way to ensure adequate code coverage in order to find malicious behaviour with these tools.
Android applications can be dynamically analysed by logging API function calls and their response to broadcast events and intents. In order to log these features from the applications at run-time, a platform is needed where an efficient approach to increase the code coverage will also be applied. Since our goal is to perform experiments to compare a random-based, state-based and a hybrid based method on real devices, we need to extract the features in all of these scenarios to make the comparison. Therefore, we applied the DynaLog dynamic analysis framework described in ~\cite{Alzaylaee2016} for all the experiments in this paper. 
\begin{figure}[t]
\centering
\includegraphics[width=3.5in,height=2.5cm]{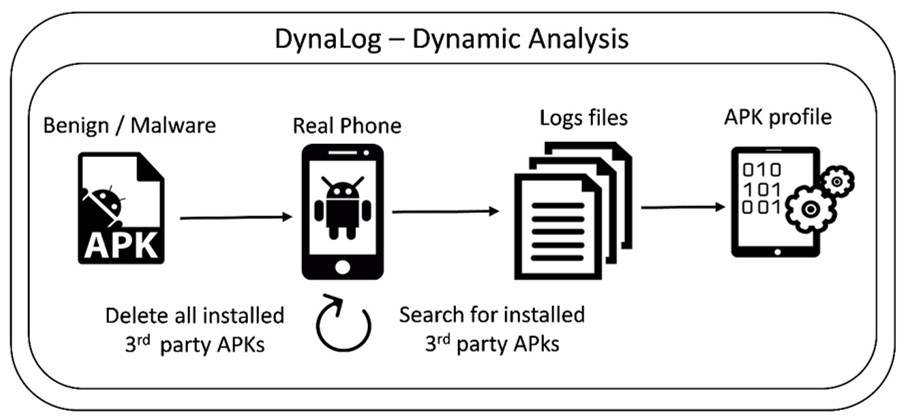}
\caption{Phone based feature extraction using DynaLog.}
\label{DynaLog}
\end{figure}

DynaLog is designed to automatically process thousands of applications, starting them sequentially in an emulator or phone, logging several dynamic features and extracting them for further processing. The framework offers the ability to instrument each application with the needed API calls to be observed, logged and extracted during the dynamic analysis. APIMonitor tool ~\cite{APIMonitor} was used to build the instrumentation component of DynaLog. It currently relies on the random-based Monkey tool for test input generation. DynaLog framework was extended to be enable device based dynamic analysis as described in ~\cite{Alzaylaee17}. From the results of the study in ~\cite{Alzaylaee17}, we utilize this extended version of DynaLog for the experiments presented in this paper using real phones to mitigate the potential impact of anti-emulation and environmental limitations of emulators on our dynamic analysis. Fig. \ref{DynaLog} shows an overview of the dynamic analysis process.

\subsection{Random input generation method}
Monkey is a random events generator provided as a part of the Android developers' toolkit, which does not require any modification to run. It is also known as an application exerciser tool that sends pseudo-random events of clicks, swipes, touch screens, gestures, etc, to a real device or an emulator. This is to ensure that most of the activities will be traversed to ensure that a high percentage of the application code is covered. In order to start Monkey, the user must to declare the number of events needed to be generated. Once this upper bound is reached, Monkey will be terminated. Monkey can be run with the Android Debug Bridge (adb) tool to test the application and report any errors that are encountered. The basic command to use monkey is as follows:
\begin{itemize}
  \item \textit{adb shell monkey \textless options\textgreater}
\end{itemize}

\begin{figure}[t]
\centering
\includegraphics[width=3.5in,height=1.8cm]{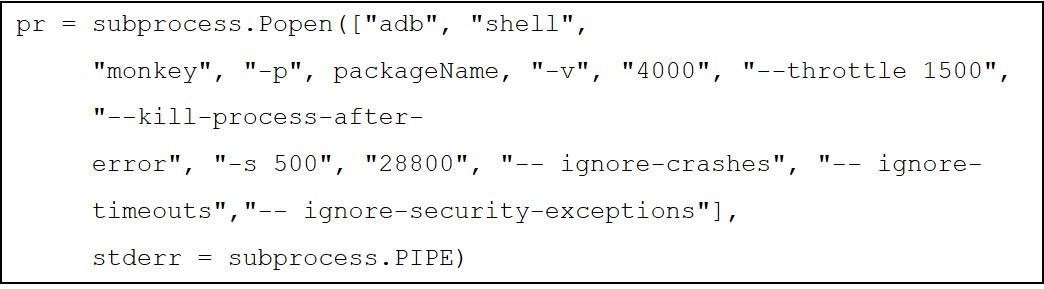}
\caption{Monkey configuration command using adb shell.}
\label{Monkey}
\end{figure}

In our experiments, we configure Monkey to run 4000 events for each application and assigned a seed value of 500 for the pseudo-random number generator. The configuration was set to ignore any security exceptions, crashes, or system timeout during the analysis (see Fig. \ref{Monkey}). Monkey continues to send events to the application UI until the number of events chosen is reached or until it encounters an error and 'crashes'. From our experience, some of the drawbacks of using Monkey on device based dynamic analysis include:

\begin{itemize}
\item The random events generated could lead to turning off the Wi-Fi connection on the phone during the analysis. The Wi-Fi would then remain turned off unless by chance another sequence of random events turns it back on (which is very rare). This situation may hide malicious behaviours that require Internet connection before they can be triggered.
\item The random events generated could also turn on the Airplane Mode thus preventing the device from sending or receiving calls, text messages and from connecting to the Internet.
\item Monkey could also generate random events that leads to unintended re-configurations of the device and some unwanted behaviour that could affect the analysis. For example, turning off adb debugging from the developer options on the phone which automatically disconnects the phone from the USB connection during analysis.  
\item Monkey can only generate UI events but not system events ~\cite{dynodroid}.
\item As it is a random-based input generation tool, Monkey often generates non-relevant events to the current state. These redundant events have no consistent pattern and cannot keep track of events that have already been covered. 
\end{itemize}

\begin{figure}[t]
\centering
\includegraphics[width=3.5in,height=1.8cm]{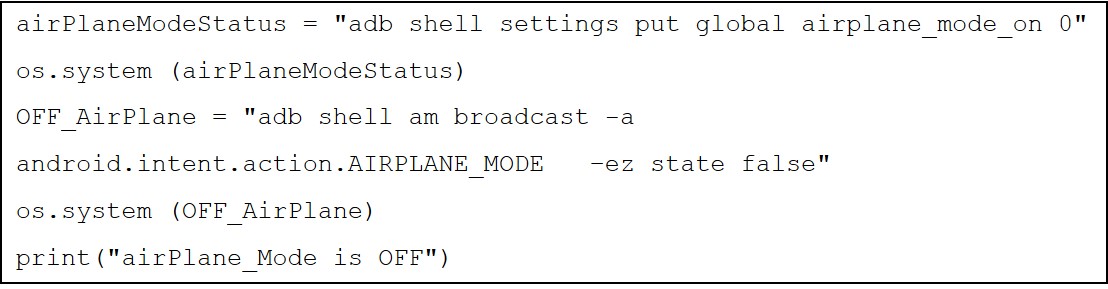}
\caption{Python code to check and restore the Airplane mode in the real phone.}
\label{airplane}
\end{figure}

\begin{figure}[b]
\centering
\includegraphics[width=3.5in,height=3.7cm]{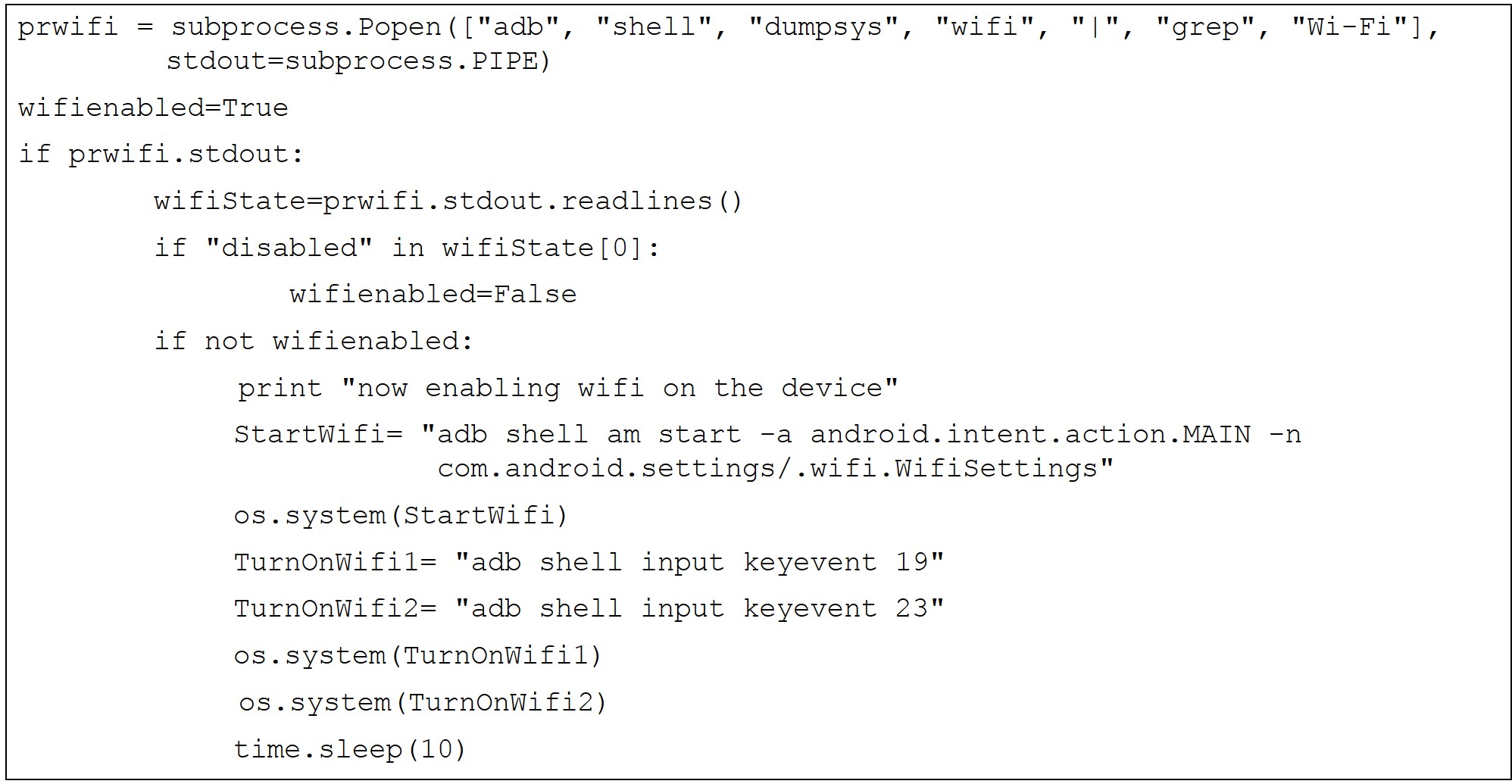}
\caption{Python code to check the Wi-Fi status and restore with the connection in the real phone.}
\label{WiFi}
\end{figure}

To overcome some of the above listed shortcomings of using Monkey in our experiments, we check the status of the Wi-Fi as well as the Airplane mode in order to ensure that they are kept in the desired mode as much as possible throughout the analyses. Fig. \ref{airplane} Illustrates the code for Airplane mode checking while Fig. \ref{WiFi} shows the code used for checking and restoring the device's Wi-Fi connectivity. The Wi-Fi status is checked prior to starting the analysis of each application and is enabled if it is currently in the disabled state. This capability was implemented using “dumpsys”, within an adb shell as illustrated in Fig. \ref{WiFi}. Two of the keyevents i.e. '19' and '23' are sent via adb shell to turn on the Wi-Fi if it is switched off. Note that, in some devices the required keyevent combination could be '20' and '23' instead of '19' and '20'.

\subsection{State-based input generation method}
In this study, we used an open source automated test input generation tool known as DroidBot ~\cite{droidbot} in order to implement a state-based approach. DroidBot is also an Android app exerciser that is considerably smarter than Monkey and can utilize static information extracted from the APK file (e.g. list of sensitive user events) that could be used to improve the event generation. It can avoid redundant re-entry of explored UI states by dynamic UI analysis. Also, DroidBot does not send random gestures (touch, click etc.) like monkey but sends specific gestures according to the position and type of the UI element (unless a 'random' event generation policy is selected). Because of this capability we refer to DroidBot as a 'state-based' generation tool.
DroidBot follows four distinct steps when testing an app:
\begin{itemize}
\item Connect to a device i.e. emulator or real device.
\item Statically analyse the apk using Androguard to infer the list of broadcasts the app may handle and the list of permission the app requires.
\item Set up the environment with device usage data like SMS logs, call logs, etc. This is set up according to 'environment policies'. 
\item Send user events. The events include gestures, broadcasts, key presses, etc., just like the events generated when a user uses the device. Same as setting up environments, DroidBot has multiple policy choices for sending events. For example, static policy is to send app-specific events according to static analysis, and dynamic policy is an extension of the static policy which improves UI event efficiency by dynamically monitoring the UI states. Since we are interested in using DroidBot as a state-based input generator, we chose the 'dynamic' policy configuration during our experiments. 
\end{itemize}

DroidBot also suffers from some limitations.  The dynamic policy mode is quite slow as the event state needs to be inferred. Monkey can generate events at a far more rapid pace with the random approach. Even though DroidBot is smarter than Monkey, it is still far less intelligent than manual testing. DroidBot is also unable to deal with unexpected situations, and can get confused by login screens or gets stuck at popup windows and drop-down menus. Since it is not feasible to incorporate manual testing in our automated dynamic analysis of thousands of applications, DroidBot's dynamic approach is perhaps the closest we can get to manual interaction. Since both random-based and the state-based methods have strengths and drawbacks, we want to obtain the benefits of both by combining their strengths in a hybrid test input generation approach.

\subsection{Hybrid input test method}
\begin{figure}[!h]
\centering
\includegraphics[width=3in,height=4.8cm]{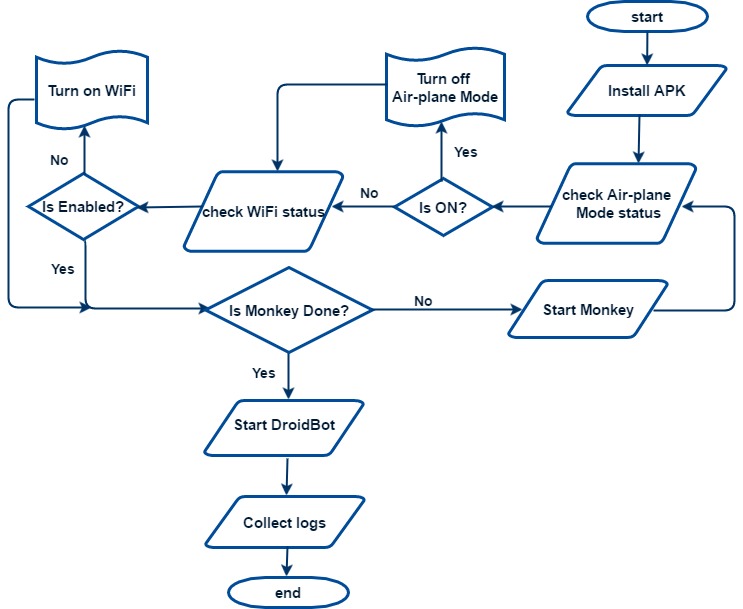}
\caption{Checking and restoring device configurations on the hybrid input generation system.}
\label{hybrid}
\end{figure}

In this subsection, we discuss our implementation of the hybrid test input generation strategy within the extended DynaLog dynamic analysis framework. As mentioned earlier, DynaLog already relies on Monkey, thus, we need to find a way to integrate DroidBot into the platform. We did this by incorporating a call to invoke a DroidBot instance after Monkey finishes sending its events. Because DroidBot closes the connection with the adb after it finishes, it would not be ideal to start DroidBot before running Monkey.
After a number of preliminary tests, we found that Monkey turns on the Airplane modes quite frequently on the devices. This is because the menu was quite easy to reach after a few touches and presses. We also found that the WiFi connection was sometimes turned off as well. Even though we implemented a check for the Airplane mode status at the beginning of each analysis, it will not turn on the Wi-Fi if has been turned off. Therefore, at the start of each analysis, the Wi-Fi status will be checked as well to make sure it is enabled. These processes will be repeated before and after the running Monkey in order to ensure that the right configuration settings are enabled before DroidBot is started. The process of checking and re-setting the device configurations during the analysis with the hybrid test input generation is illustrated in Fig. \ref{hybrid}.

\section{Methodology and Experiments}
\subsection{Testbed Setup}
The experiments to evaluate the impact of the  hybrid test input generation on the dynamic analysis of Android apps   were performed using real phones with the following configurations. The phones were: an Elephone P9000 brand equipped with Android 6.0 Marshmallow, Helio P10 MTK6755 Octa Core CPU, 4GB RAM, 4G LTE, 32 GB of ROM, and 32 GB of external SD card storage. Moreover, each phone was installed with a sim card that has some credit to enable sending SMS, outgoing calls, and 3G data usage. The phones also were connected to an internal Wi-Fi access point with internet connection to allow the analysed apps to communicate with their external servers if needed.  The analysis was performed in three scenarios. First, all the apps were processed and analysed using Monkey only. Then, they were run again in the analysis environment using DroidBot. Lastly, the system with input test generation from both Monkey and DroidBot together was used to accomplish the hybrid input test generation in the third scenario.

\subsection{App Feature Extraction}
Once all the applications are run within the three scenarios, the logs are collected in text files which are further processed into a single .csv file for each scenario. The .csv files contains '0's and '1's symbolizing the presence or absence of each feature being extracted from the log files. 

\subsection{Dataset}
We have analysed 2444 Android samples in order to extract benign and malware apps' features. Of these, 1222 benign samples were obtained from McAfee Labs (Intel Security). Our malware samples consists of 1222 which belong to 49 Android malware families of the Android malware genome project ~\cite{malgenomeproject}.  

\section{Results and Discussions}
This section presents the results of the experiments performed to compare the novel hybrid input test generation method with the random-based and state-based standalone methods. 
\subsection{Experiment 1: Monkey vs. Hybrid}
In order to evaluate the efficiency of the hybrid-based test input generation method, we analysed the extracted features from the scenario with Monkey alone and the scenario with  hybrid generation. Table I and Table II show the top-10 extracted features from malware and benign dataset respectively using the hybrid test input compared to the use of Monkey. Both tables show that more features are able to be extracted from the hybrid-based analysis compared to those from the Monkey-based analysis using the same application set. 
The API method call \textit{“Ljava/io/File;-\textgreater exists”}, for instance, was logged from 677 malware APKs using the hybrid test input, while it was only logged from 477 malware APKs using the Monkey-based generation. By the same token, the method \textit{“Landroid/telephony/TelephonyManager;-\textgreater getDeviceId”} in Table I, has been extracted from 429 malware APKs using the hybrid method, whereas only 315 malware APKs logged the same API method when Monkey was used. 

Similar discoveries appear with the benign samples with even higher differences observed as shown in Table II. With some of the API calls, the difference between the hybrid and Monkey test input generation were \textgreater 200. For example, the method \textit{“Landroid/net/Uri;-\textgreater parse”} was logged from 492 benign APKs using the hybrid method, while the same method was extracted from only 192 benign APKs using Monkey. With dynamic analysis, the malware detection mechanism will likely perform better with more extracted API call features. Overall, the hybrid test input generation shows a much higher efficiency of extracting and logging API calls for the Android applications during the run-time analysis.

\begin{table}[t]
\centering
\tiny
\caption{Top-10 API calls logged from malware samples using the hybrid method compared to monkey}
\label{my-label}
\begin{tabular}{|p{4.5cm}||c|c|c|}
\hline
\multicolumn{1}{|c|}{\textbf{API signatures}}            & \textbf{Monkey} & \textbf{Hybrid} & \textbf{Difference} \\ \hline
Ljava/io/File;-\textgreater exists                                   & 477             & 667             & 190                 \\ \hline
Ljava/security/MessageDigest;-\textgreater getInstance               & 338             & 473             & 135                 \\ \hline
Landroid/content/pm/ApplicationInfo;-\textgreater getApplicationInfo & 435             & 563             & 128                 \\ \hline
Ljava/security/MessageDigest;-\textgreater digest                    & 310             & 431             & 121                 \\ \hline
Ljava/util/zip/ZipInputStream;-\textgreater read                    & 219             & 336             & 117                 \\ \hline
Landroid/telephony/TelephonyManager;-\textgreater getDeviceId      & 315             & 429             & 114                 \\ \hline
Ljava/util/TimerTask;-\textgreater \textless init\textgreater                            & 808             & 921             & 113                 \\ \hline
Landroid/content/pm/PackageManager                       & 651             & 756             & 105                 \\ \hline
Lorg/apache/http/client/HttpClient;-\textgreater execute             & 89              & 188             & 99                  \\ \hline
Ljava/io/File;-\textgreater mkdir                                    & 274             & 366             & 92                  \\ \hline
\end{tabular}
\end{table}

\begin{table}[t]
\centering
\tiny
\caption{Top-10 API calls logged from benign samples using the hybrid method compared to monkey}
\label{my-label}
\begin{tabular}{|p{4.5cm}||c|c|c|}
\hline
\multicolumn{1}{|c|}{\textbf{API Signatures}}        & \textbf{Monkey} & \textbf{Hybrid} & \textbf{Difference} \\ \hline
Landroid/net/Uri;->parse                             & 192             & 492             & 300                 \\ \hline
Ljava/util/zip/ZipInputStream;-\textgreater read                 & 173             & 408             & 235                 \\ \hline
Ljava/security/MessageDigest;-\textgreater digest                & 297             & 519             & 222                 \\ \hline
Ljava/security/MessageDigest;-\textgreater getInstance           & 327             & 549             & 222                 \\ \hline
Lorg/apache/http/client/HttpClient;-\textgreater execute         & 56              & 264             & 208                 \\ \hline
Ljava/lang/reflect/Method;-\textgreater getClass                 & 359             & 565             & 206                 \\ \hline
Ljava/lang/Class;-\textgreater getName                           & 217             & 422             & 205                 \\ \hline
Landroid/content/Context;-\textgreater getResources              & 437             & 639             & 202                 \\ \hline
Ljava/util/TimerTask;-\textgreater \textless init\textgreater                        & 575             & 774             & 199                 \\ \hline
Landroid/content/pm/PackageManager;-\textgreater checkPermission & 164             & 359             & 195                 \\ \hline
\end{tabular}
\end{table}

\begin{table}[t]
\centering
\tiny
\caption{Top-10 API calls logged from malware samples using the hybrid method compared to DroidBot}
\label{my-label}
\begin{tabular}{|p{4.5cm}||c|c|c|}
\hline
\multicolumn{1}{|c|}{\textbf{API signatures}} & \textbf{DroidBot} & \textbf{Hybrid} & \textbf{Difference} \\ \hline
Ljava/util/Date                               & 177               & 301             & 124                 \\ \hline
Ljava/util/Date;-\textgreater \textless init\textgreater                      & 171               & 289             & 118                 \\ \hline
Ljava/util/List                               & 171               & 289             & 118                 \\ \hline
Ljava/util/Timer;-\textgreater schedule                   & 222               & 339             & 117                 \\ \hline
Ljava/util/GregorianCalendar;-\textgreater getTime        & 107               & 207             & 100                 \\ \hline
Ljava/util/zip/ZipInputStream;-\textgreater read          & 242               & 336             & 94                  \\ \hline
Ljava/io/File;-\textgreater exists                        & 602               & 667             & 65                  \\ \hline
Ljava/security/MessageDigest;-\textgreater digest         & 366               & 431             & 65                  \\ \hline
Lorg/apache/http/client/HttpClient;-\textgreater execute  & 133               & 188             & 55                  \\ \hline
Ljava/security/MessageDigest;-\textgreater update         & 288               & 335             & 47                  \\ \hline
\end{tabular}
\end{table}

\begin{table}[t]
\centering
\tiny
\caption{Top-10 API calls logged from benign samples using the hybrid method compared to DroidBot}
\label{my-label}
\begin{tabular}{|p{4.5cm}||c|c|c|}
\hline
\multicolumn{1}{|c|}{\textbf{API signatures}}        & \textbf{DroidBot} & \textbf{Hybrid} & \textbf{Difference} \\ \hline
Landroid/os/Process;-\textgreater myPid                          & 87                & 106             & 19                  \\ \hline
Landroid/net/Uri;-\textgreater parse                             & 480               & 492             & 12                  \\ \hline
Landroid/media/AudioManager;-\textgreater getStreamVolume        & 37                & 47              & 10                  \\ \hline
Landroid/content/res/AssetManager;-\textgreater open             & 320               & 328             & 8                   \\ \hline
Landroid/net/NetworkInfo;-\textgreater getExtraInfo              & 23                & 30              & 7                   \\ \hline
Ljava/lang/reflect/Method;-\textgreater getClass                 & 559               & 565             & 6                   \\ \hline
Landroid/telephony/TelephonyManager;-\textgreater getSimOperator & 72                & 77              & 5                   \\ \hline
Ljava/lang/reflect/Method;-\textgreater getMethod                & 245               & 249             & 4                   \\ \hline
Ljava/util/GregorianCalendar;-\textgreater getTime               & 188               & 191             & 3                   \\ \hline
Landroid/content/Context;-\textgreater getAssets                 & 243               & 246             & 3                   \\ \hline
\end{tabular}
\end{table}

\subsection{Experiment 2: DroidBot vs. Hybrid}
Table III and Table IV present the top-10 features extracted from both malware and benign samples respectively using the hybrid method vs. DroidBot. With the malware set as in Table III, it can be seen clearly that the hybrid generation allows for the discovery of more API calls with a difference of over 100 in some cases. For instance, the class \textit{“Ljava/util/Date”} was extracted from only 177 malware samples using DroidBot, while with the hybrid method, it was logged from 124 more malware samples. The differences decreased to less than 20 samples when we applied the same analysis to the benign sample set. Which indicates that DroidBot is far intelligent than Monkey but still is not as efficient as the hybrid method.

\subsection{Experiment 3: Monkey vs. DroidBot}
\begin{figure}[t]
\centering
\includegraphics[width=3.5in,height=3.5cm] {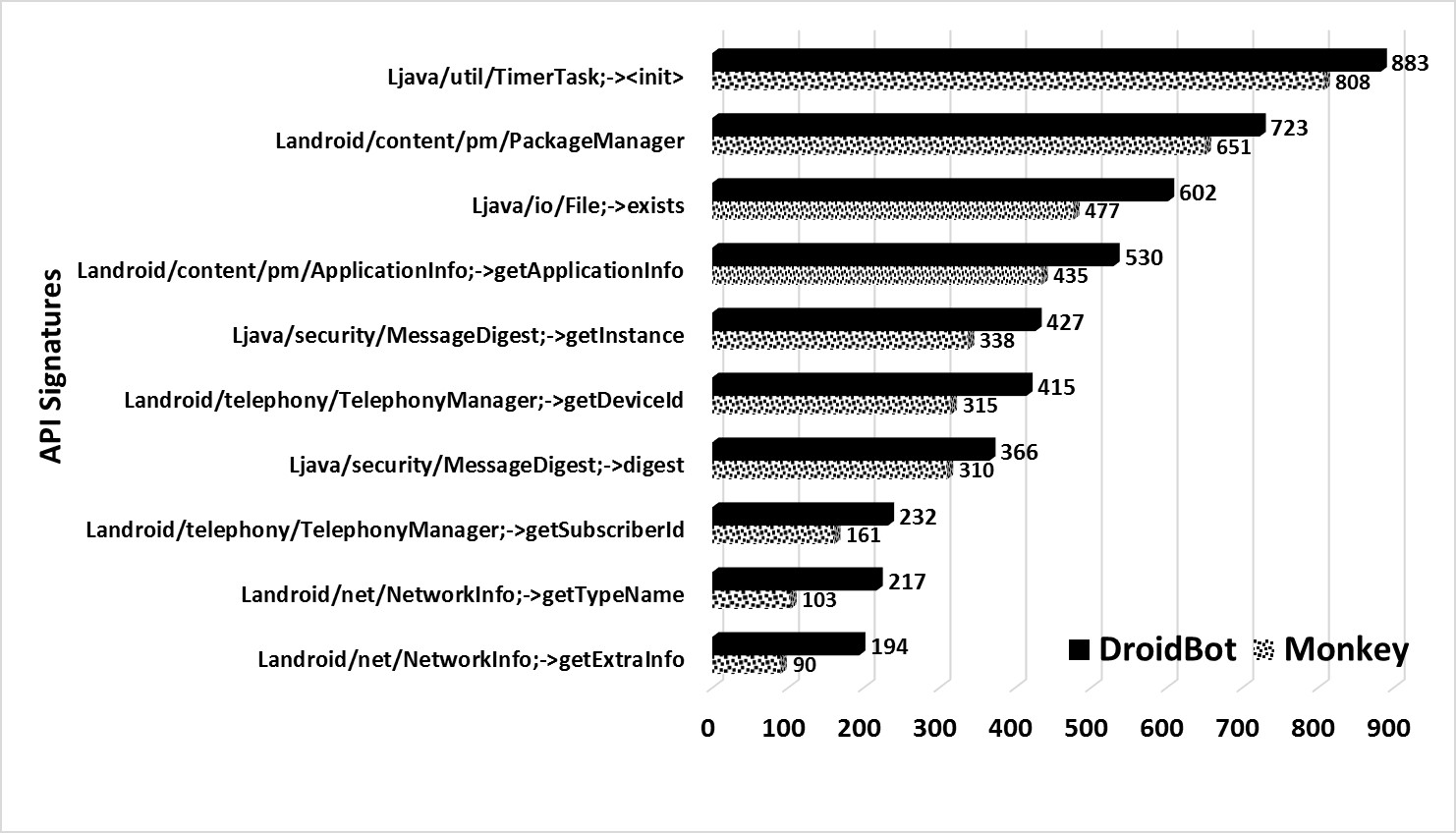}
\caption{Top-10 logged API calls from the malware set where DroidBot was  better than Monkey.}
\label{DroidBot-malware}
\end{figure}

In this subsection, we compare the results from Monkey based analysis to the DroidBot based analysis. In this experiment, we discovered that some API calls were logged at a higher number with Monkey whereas others were logged at a higher number with DroidBot, using the same sample set. Fig. \ref{DroidBot-malware} and Fig. \ref{Monkey-better} show clearly the differences between Monkey and DroidBot. Fig. \ref{DroidBot-malware} illustrates the differences of the extracted features from the malware samples where DroidBot outperforms Monkey. The differences sometimes exceeds  100 for some features extracted from the Malware sample set using DroidBot vs. Monkey. The differences are even higher with the benign sample set. Fig. 7 show the top-10 API calls where DroidBot exceeded Monkey with the benign samples. For example,  \textit{“Landroid/net/Uri;-\textgreater parse”} was extracted from 480 benign APKs using DroidBot compared to only 192 using Monkey. 

In a few exceptional cases, Monkey was better than DroidBot (as shown in Fig. \ref{Monkey-better}) with the malware sample set. More than 100 malware samples logged  \textit{“Ljava/util/Date;-\textgreater \textless init\textgreater”} using Monkey compare to  DroidBot (Fig. \ref{Monkey-better}). This indicates that Monkey, with its random approach, could sometimes reach events that DroidBot may be unable to.

Consequently, we can conclude that even though DroidBot appears to perform better than Monkey in the overall analysis, since in some cases Monkey logs API calls that DroidBot could not, both tools can effectively complement each other. Hence, combining both of them in a hybrid approach makes sense in order to improve code coverage. 

Due to the lack of space, we present comparative results for methods form the TelephonyManager class which most malicious apps have been reported to commonly utilize (Fig. \ref{TelephonyManager}). The figure shows the overall number of collected API call logs form the three scenarios using the malware sample set. From the figure, it can be seen that the hybrid methods enables more of extraction of the API calls, followed by DroidBot, then Monkey with the lowest performance.

Fig. \ref{overall-malware} and Fig. \ref{overall-benign} show graphs of comparative performance of our proposed hybrid method vs. DroidBot and Monkey for malware and benign samples. Our proposed method exceeded the other two examined input test generation methods in most of the logged API signatures. However, we observed that the overall difference between the hybrid and DroidBot performance is larger in the malware set compared to the benign set as can be seen in the graphs.
 
From the malware sample set 76 API calls showed differences in numbers between the three methods. The summary of overall findings from the malware sample set are as follows: 
\begin{itemize}
\item Out of the 76 API calls with differences, 64 of these showed the proposed Hybrid method with higher logs than DroidBot.
\item Out of the 76 API calls with differences, 62 showed the proposed Hybrid method having higher logs than Monkey.
\item Out of the 76 API calls with differences, 49 showed DroidBot having higher logs than Monkey.
\item Out of the 76 API calls with differences, 23 showed Monkey having higher logs than DroidBot.
\item Out of the 76 API calls with differences, 10 showed Monkey with higher logs than the proposed Hybrid method.
\item Out of the 76 API calls with differences, 4 have DroidBot showing higher logs than the proposed Hybrid method. 
\end{itemize}

From the above summary we can conclude that the hybrid test input generation method surpassed Monkey and the DroidBot methods in overall extraction of API calls from both malware and benign Android applications.

\begin{figure}[b]
\centering
\includegraphics[width=3.5in,height=3.5cm]{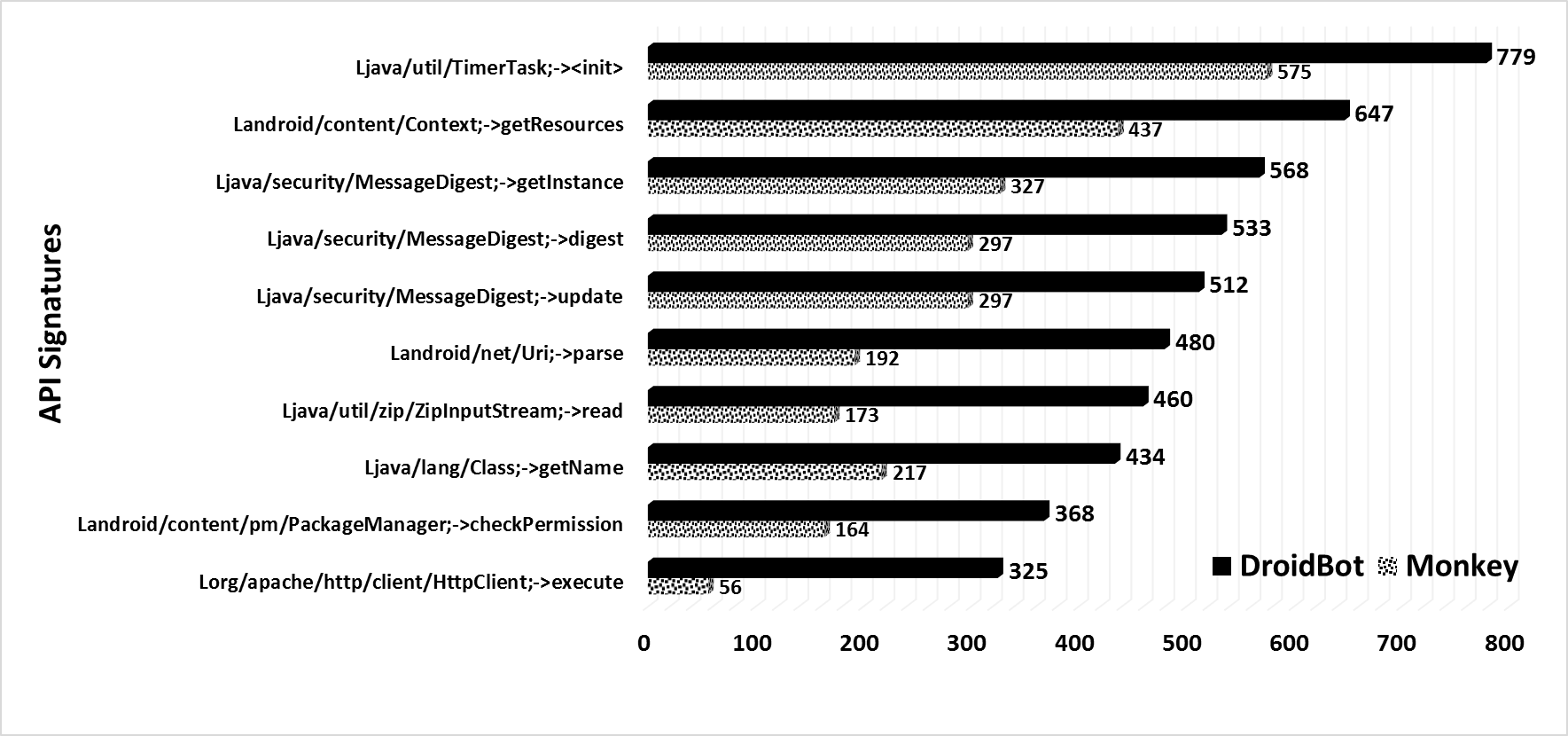}
\caption{Top-10 logged API calls from the benign set where DroidBot was better than Monkey.}
\label{DroidBot_benign}
\end{figure}

\begin{figure}[t]
\centering
\includegraphics[width=3.5in, height=3.5cm]{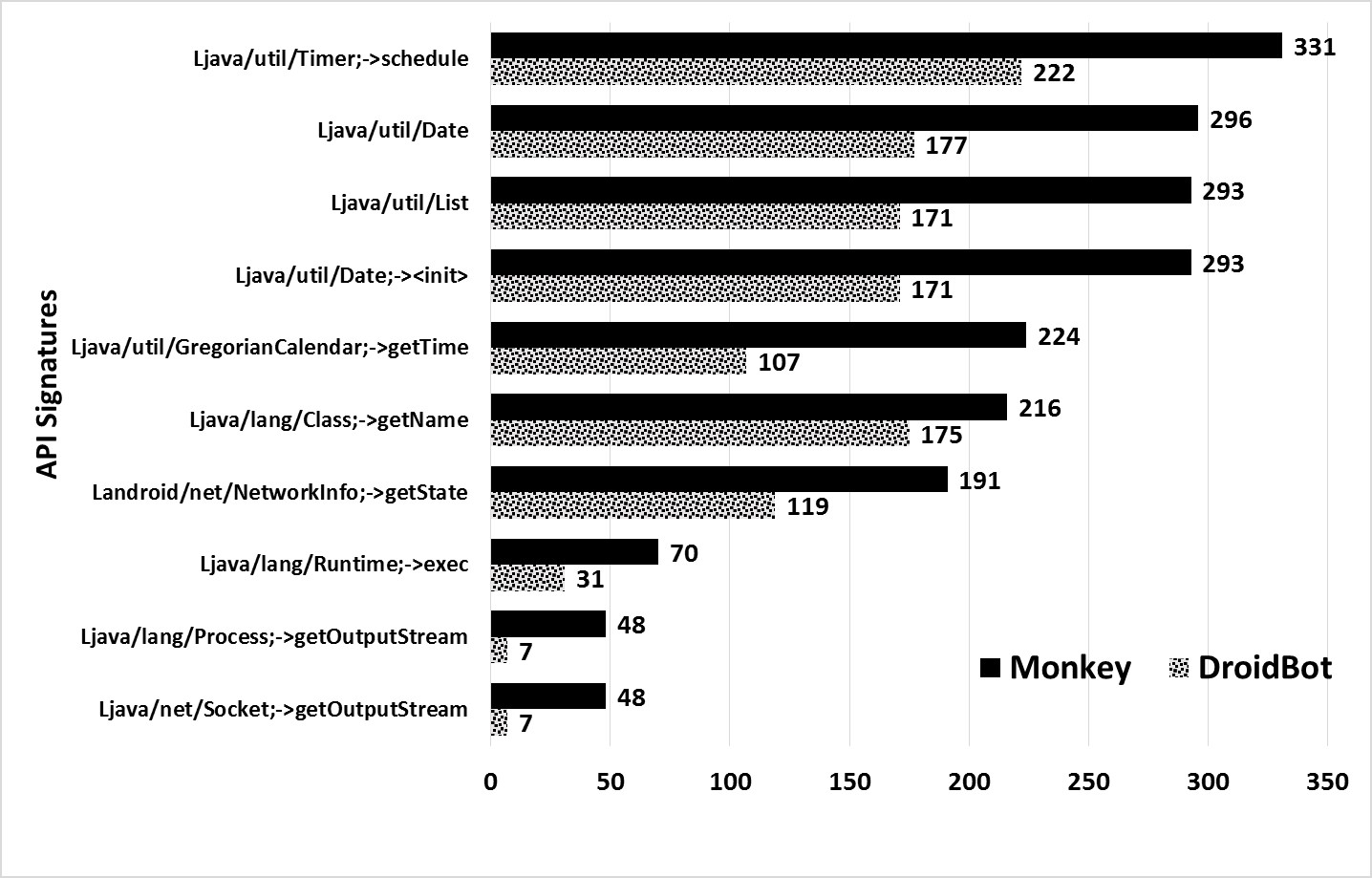}
\caption{Top-10 logged API calls from the malware set where Monkey was better than better than DroidBot.}
\label{Monkey-better}
\end{figure}

\begin{figure}[b]
\centering
\includegraphics[width=3.5in,height=3.5cm]{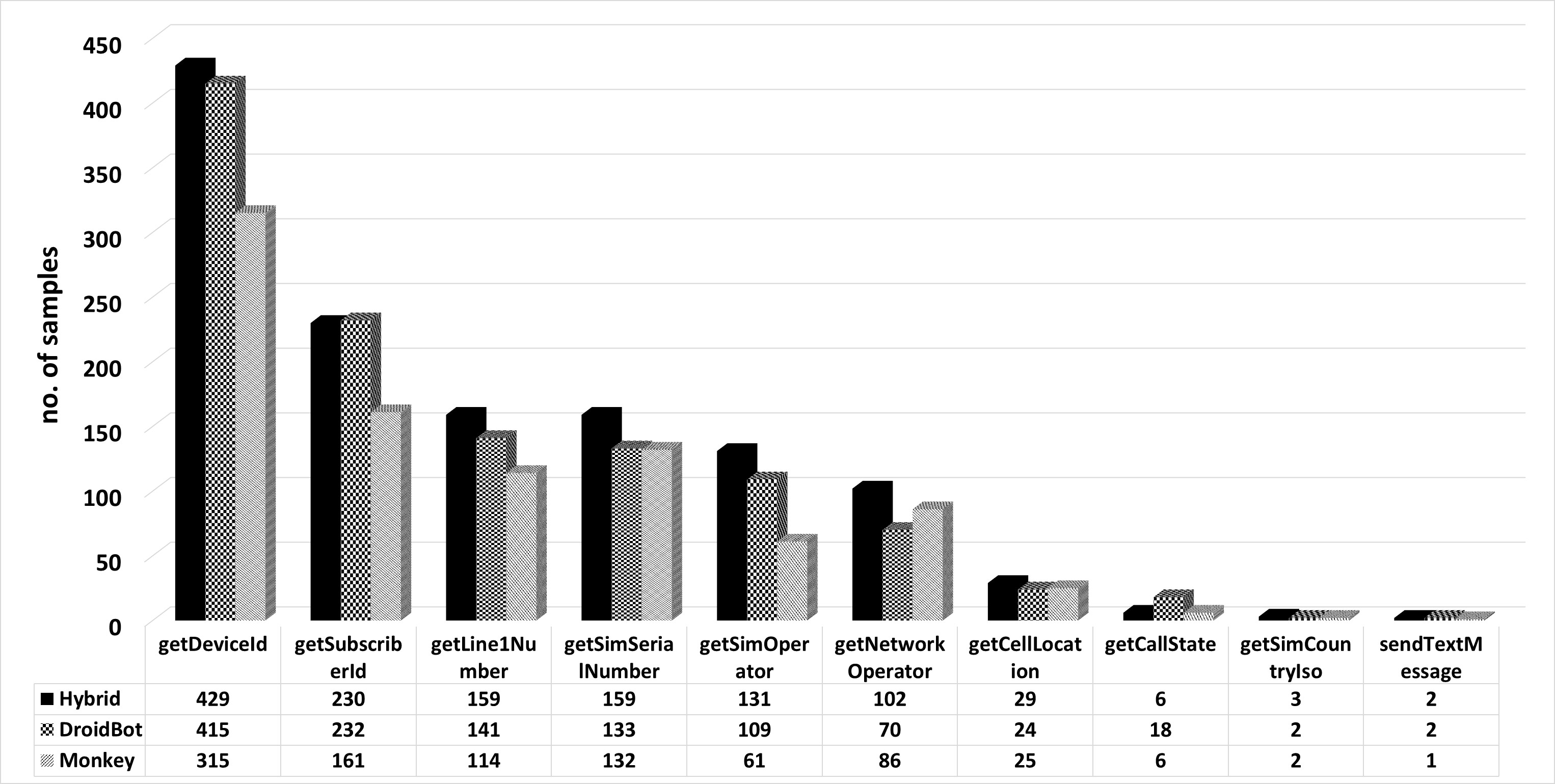}
\caption{The overall comparisons of number of the logged methods from TelephonyManager API class using malware sample test.}
\label{TelephonyManager}
\end{figure}

\begin{figure*}
\centering
\includegraphics[width=7in,height=8cm]{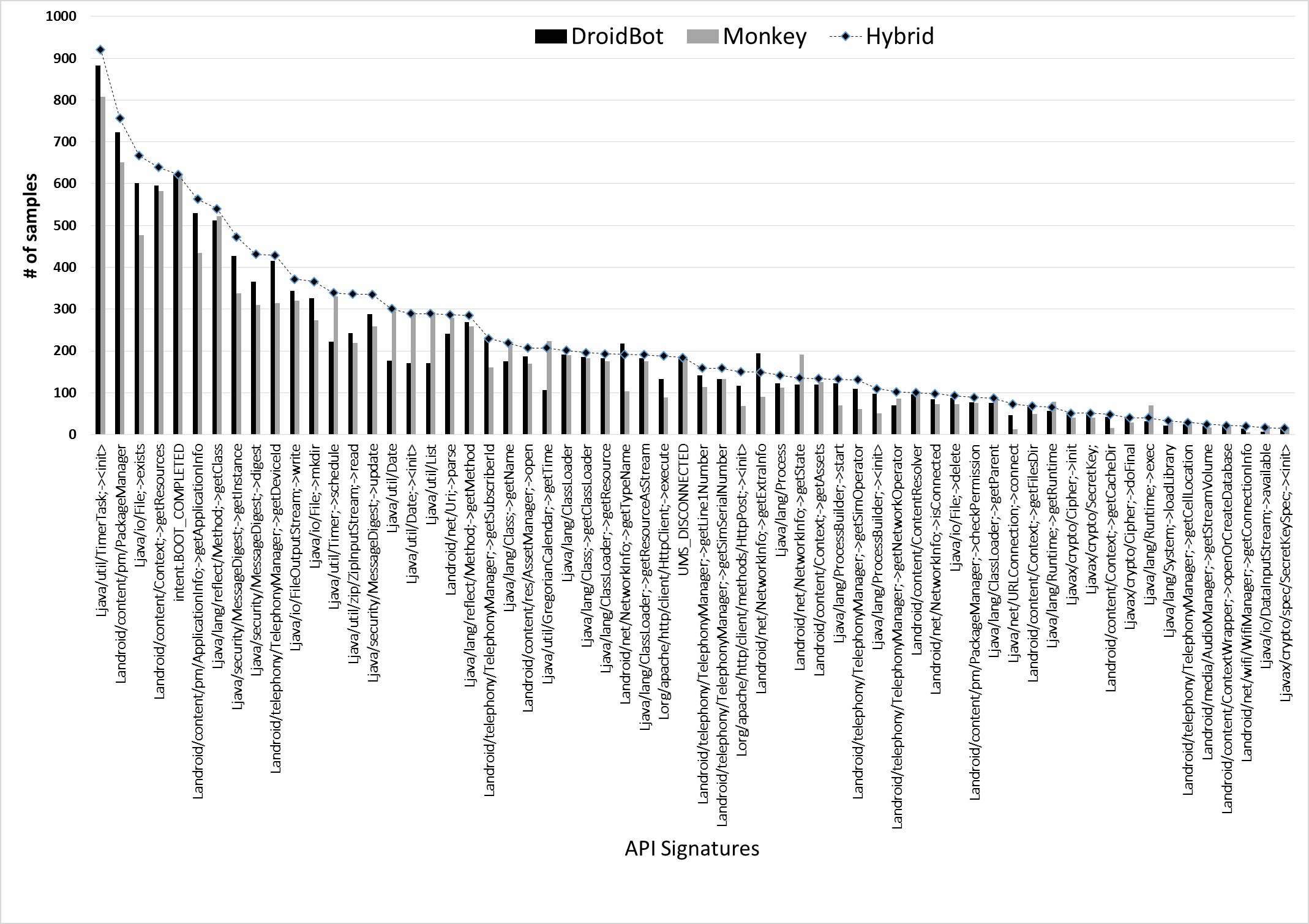}%
\hfil
\caption{DroidBot vs. Monkey vs. Hybrid: Number of APKs where logs of the given API calls were found in the malware set.}
\label{overall-malware}
\end{figure*}

\begin{figure*}
\centering
\includegraphics[width=7in,height=8cm]{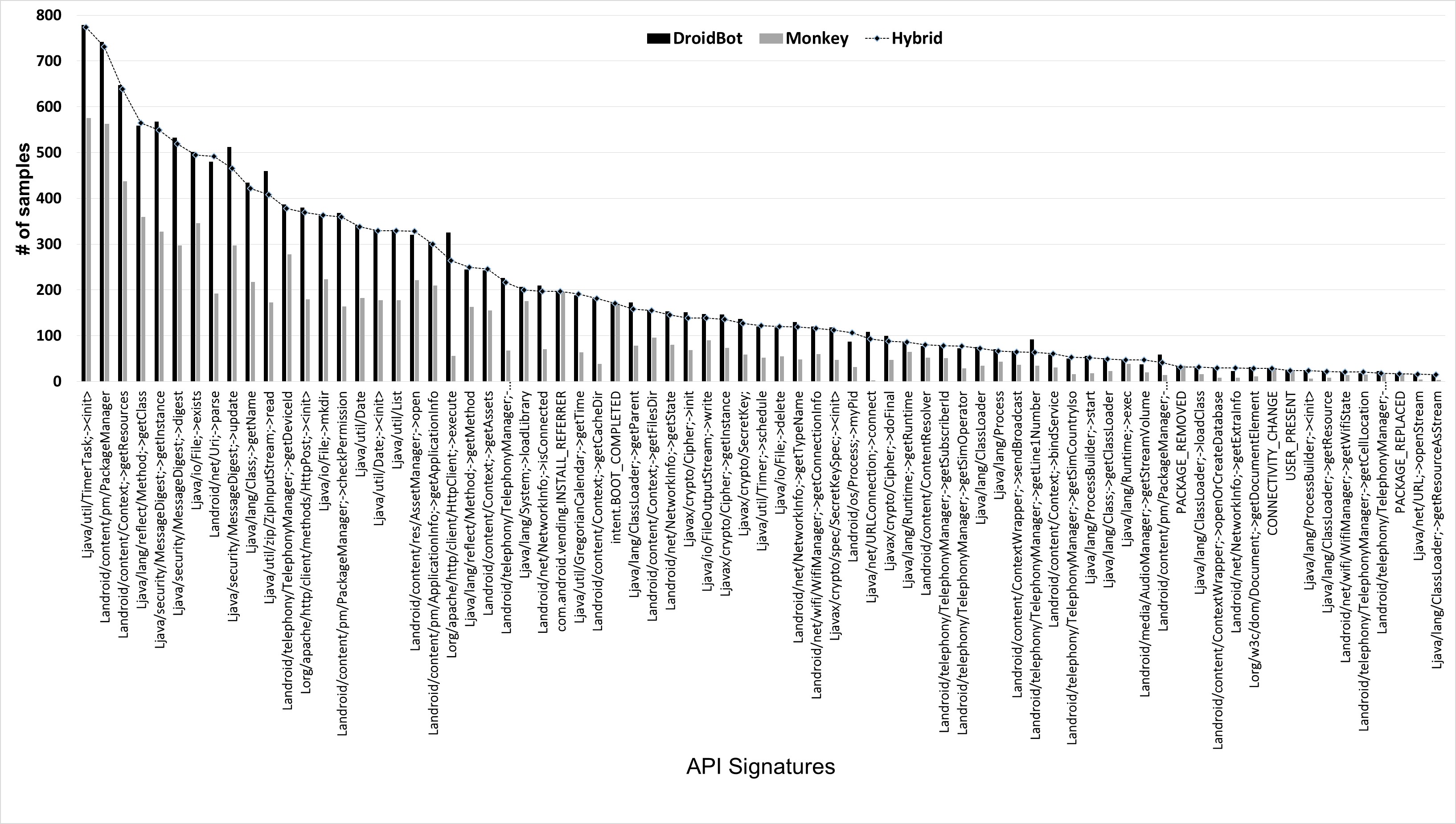}%
\hfil
\caption{DroidBot vs. Monkey vs. Hybrid: Number of APKs where logs of the given API calls were found in the benign set.}
\label{overall-benign}
\end{figure*}

\section{Related Work}

Dynamic analysis has been growing in popularity in the field of Android malware detection. Several efforts have  been made towards improving the effectiveness of applying Dynamic analysis to detect Android malware. Dynamic analysis uses predefined scripts that will be executed while the application is running in an emulator or real device in order to collect several  behaviour indicators. Code coverage is an important factor in dynamic analysis because it determines how much of the malicious behaviour can be discovered. Hence, input generation needs to be as efficient as possible. Contrary to previous work on input test generation approaches or dynamic analysis of Android apps, we attempt to improve the code coverage efficiency by combining a random method using Monkey with state-based method using DroidBot and evaluate the method using real phones. 
Sophisticated Android malware employ techniques such as reflection, code obfuscation, dynamic loading, and native code which may hinder static analysis tools ~\cite{flowdroid}, ~\cite{gorla2014checking}. Thus, to overcome limitations of static analysis researchers tend to utilize dynamic analysis to investigate the apps during execution. The most popular input test generation tool used in most dynamic analysis systems is Monkey.  Monkey ~\cite{monkey} is a command-line tool which developers can configure to run on an emulator or real device in order to generate pseudo-random streams of user events such as touches, gestures, or clicks. It is popular due to the simplicity of its configuration and use and is also readily available as part of the Android developers' toolkit. However, the random testing approach may not be so effective in triggering Android malware behaviours.
Dynodroid ~\cite{dynodroid} is a random exploration tool that generates both UI inputs and system inputs to Android applications which is based on the principle of “observe-select-execute”.  It also allows combining inputs from machine and human. However, it requires instrumentation of the Android framework in order to generate the system events. It can also run only on an emulator, whereas malware applications can implement anti-emulation techniques in order to avoid detection. Dynodroid restricts the apps from communicating with other apps which may affect the analysis of the malicious apps. 
A3E ~\cite{azim2013targeted} is publicly available tool that consists of two strategies to trigger the applications: the DFS (depth first search) and a taint-targeted approach. However, the open source A3E repository does not provide the taint-targeted strategy. ACTEve ~\cite{anand2012automated} is developed to support both system events and UI events which is based on concolic-testing and symbolic execution. However, ACTEve needs to instrument both the Android framework and the application in order to perform the test. 
An empirical study that evaluates several test input generation tools  was done by ~\cite{automated}. It comparatively evaluates Monkey ~\cite{monkey}, Dynodroid ~\cite{dynodroid}, ACTEve ~\cite{anand2012automated}, A3E ~\cite{azim2013targeted}, GUIRipper ~\cite{amalfitano2012using}, SwiftHand ~\cite{choi2013guided}, and PUMA ~\cite{puma}, in terms of the code coverage. The study reveals that among the tested tools, Monkey performs best on average considering the four measurement metrics used in the study. However, this test was conducted on emulators unlike our work which was performed on real devices. Moreover, this study did not include DroidBot which we used in our study. DroidBot meets our requirements more than other tools because it could run on devices and does not require instrumentation to the Android framework. Unlike the previous works, this paper presents a comparative analysis of a proposed hybrid method that utilizes a combination of random-based and state-based input test generation methods. This is to ensure that more malicious behaviour can be reached and triggered as much as possible due to potentially higher code coverage.

\section{Conclusion}
In this paper, we presented a novel hybrid test input generation approach that enables us to trigger more malicious behaviours of  Android malware during dynamic analysis. We designed and implemented the hybrid strategy to combine a random-based with a state-based input test generation method and incorporated this within a framework that enables automated mass dynamic analysis of applications on real devices. By doing so we are able to leverage the advantages of the random mechanism along with those of the state-based technique to trigger more events and increase the code coverage and hence the possibility of logging more malicious behaviour indicated by API function calls. We have performed several experiments to comparatively evaluate the hybrid strategy with the random and state-based method on real devices and the results shows its capability to trigger more malicious behaviours. The results also showed that several API calls were extracted more effectively from the hybrid test input generation than the single based generation. Thus, we conclude that the use of the hybrid test input generation would improve dynamic analysis code coverage and potentially impact the detection of Android malware. For future work, we propose to evaluate the performance of the hybrid tool using larger sample datasets. Additionally, we intend to evaluate the impact on detection systems that can utilize the API calls collected using the hybrid method for detection of Android malware.






\bibliographystyle{IEEEtran}
\bibliography{bibliography}
%



\end{document}